

\input phyzzx

 \def\12{1\over2}

 \tolerance=500
 \overfullrule=0pt
 \Pubnum={\vbox { \hbox{CERN-TH.6243/91}\hbox{UGVA-PHY-09-746/91}}}
 \pubnum={CERN-TH.6243/91}
 \date={September, 1991}
 \pubtype={}
 \titlepage

\def\const{{ \rm const. }}

\def\alm{{ \alpha_- }}
\def\alp{{ \alpha_+ }}
\def\bem{{ \beta_- }}
\def\bep{{ \beta_+ }}

\def\rom{{ {\rho-1 \over 2} }}
\def\rop{{ {\rho+1 \over 2} }}
\def\ropm{{ {\rho'-1 \over 2} }}
\def\ropp{{ {\rho'+1 \over 2} }}

 \title{SOME CORRELATION FUNCTIONS OF MINIMAL SUPERCONFORMAL MODELS
COUPLED TO SUPERGRAVITY}
 \author{L. Alvarez-Gaum\'e}
 \address{Theory Division, CERN\break CH-1211 Geneva 23, Switzerland}
\vskip 2cm
\centerline{and}
\author{Ph. Zaugg}\foot{Partially supported by
the Swiss National Science Foundation.}

\address{D\'epartement de Physique Th\'eorique\break Universit\'e de
Gen\`eve\break CH-1211 Gen\`eve 4, Switzerland}

\abstract{We compute general three-point functions of
minimal superconformal models coupled to supergravity in the
Neveu-Schwarz sector for spherical
topology thus extending to the superconformal case the results of
Goulian and Li and  of Dotsenko.}
 \endpage
 \pagenumber=1

\def\rseiberg{N. Seiberg: {\it Notes on Quantum Liouville Theory and
Quantum Gravity}.  RU-90-29.  Talk presented at
the 1990 Yukawa Interntional
Seminar.}

\def\rdifrankutone{P. Di Francesco and
D. Kutasov: Nucl. Phys.{\bf B342} (1990)
589.}

\def\rdotfat{ V.S. Dotsenko and
V.A. Fateev: Nucl. Phys.{\bf B240} (1984) 312;
 Nucl. Phys.{\bf B251} (1985) 691; Phys. Lett {\bf 154B} (1985) 291.}

\def\rdotsenko{V.S. Dotsenko: PAR-LPTHE 91-18.}

\def\rgoulian{M. Goulian and
M. Li: Phys. Rev. Lett. {\bf 66}-16 (1991) 2051.}

\def\rpolyakov{A.M. Polyakov: Mod. Phys. Lett. {\bf A6} (1990) 635.}

\def\rkitazawa{Y. Kitazawa: HUTP-91/A013.}

\def\rdifrankuttwo{P. Di Francesco
and D. Kutasov: Phys. Lett. {\bf 261B}
(1991) 385.}

\def\rberkleb{ M. Bershadsky
and I. Klebanov: PUPT-1241 (1991).}

\def\rdoubles{E. Brezin and
 V. Kazakov : Phys. Lett. {\bf 236B} (1990) 144;
M. Douglas and
S. Shenker: Nucl. Phys.{\bf B335} (1990) 635; D.J. Gross and
A.A. Migdal: Phys. Rev. Lett. {\bf 64} (1990) 717.}

\def\rbpz{A.A. Belavin, A.M. Polyakov and A.B. Zamolodchikov:
 Nucl. Phys.{\bf B241} (1984) 333.}

\def\rpolyakov{A.M. Polyakov: Mod. Phys. Lett.{\bf A2} (1987) 893.}

\def\rkpz{V.G. Knizhnik, A.M. Polyakov and A.B. Zamolodchikov:
Mod. Phys. Lett.
{\bf A3} (1988) 819.}

\def\rpz{A.M. Polyakov,
 A.B. Zamolodchikov: Mod. Phys. Lett.{\bf A3} (1988) 819.}

\def\rdistler{J. Distler and
 H. Kawai:  Nucl. Phys.{\bf B321} (1989) 509.}

\def\rdavid{F. David: Mod. Phys. Lett. {\bf A3} (1988) 1651.}

\def\rdhk{J. DIstler,
Z. Hlousek and H. Kawai: Int. J. Mod. Phys. {\bf A5} (1990)
391.}

\def\rlagthree{L. Alvarez-Gaum\'e,
 J.L.F. Barb\'on and C. G\'omez: {\it
Fusion Rules in Two-Dimensional Gravity}.
 CERN-TH-6142/91.  To appear in Nucl. Phys.B.}

\def\rbpz{A.A. Belavin, A.M. Polyakov and A.B. Zamolodchikov:
 Nucl. Phys.{\bf B241} (1984) 333.}

\def\rfqsone{D. Friedan, Z. Qiu and
S. Shenker: In {\it Vertex Operators in Mathematical
Physics}. J. Lepowsky ed. Springer Verlag, 1984.}

\def\rfqstwo{D. Friedan, Z. Qiu and
S. Shenker: Phys. Rev. Lett. {\bf 51} (1984)
1575.}

\def\rfqsthree{D. Friedan, Z. Qiu and
S. Shenker: Phys. Lett. {\bf 151B} (1985) 37.}

\def\rbkt{M. Bershadsky,
V. Knizhnik and A. Teitelman: Phys. Lett. {\bf 151B} (1985) 31.}

\def\reich{H. Eichenherr: Phys. Lett. {\bf 151B} (1985) 26.}

\def\rqiu{Z. Qiu: Nucl. Phys. {\bf B270} (1986) 205.}

\def\rmussardo{G. Mussardo,
G. Sotkov and H. Stanishkov:  Phys. Lett. {\bf 195B} (1987) 397;
Nucl. Phys. {\bf B305} (1988) 69.}

\def\rdouglas{M. Douglas: Phys. Lett. {\bf238B} (1990) 176.}

\sequentialequations

{\bf 1}. {\it Introduction}.  The successes of
the double scaling limit
\REF\doubles{\rdoubles} [\doubles] and
its connection with the KP-hierarchy
\REF\douglas{\rdouglas}[\douglas] in
the computation of correlation functions
of minimal models coupled to $2D$-gravity (see for example
\REF\difrankutone{\rdifrankutone}
[\difrankutone]) prompted a good deal of activity
in trying to reproduce the same
 results directly in the continuum limit
(for details and references see
\REF\seiberg{\rseiberg}\REF\lagcg{L. Alvarez-Gaum\'e,
 C. Gomez: {\it Topics in
Liouville Theory}. In Proceedings of
the Trieste Spring School 1991.  R. Dijkgraaf,
S. Randjbar-Daemi and H. Verlinde eds.
 World Scientific (to appear). CERN-TH-6175/91.}
[\seiberg,\lagcg]).

  The original approach to the coupling of Conformal Field
Theories (CFT) to gravity appeared in
\REF\polyakov{\rpolyakov}
\REF\kpz{\rkpz}
[\polyakov,\kpz].  Using the light-cone gauge
these authors were able to exactly
compute the gravitational dimensions of
the gravitationally dressed primary fields.
These results were obtained subsequently in
the conformal gauge
\REF\distler{\rdistler}
\REF\david{\rdavid}[\distler,\david] which
 also allowed the generalization
of some of the results to non-spherical
topologies.  Several methods have
been suggested for the computation of correlation functions
 in Liouville theory
coupled to minimal conformal models
\REF\bpz{\rbpz}
\REF\fqstwo{\rfqstwo}[\bpz,\fqstwo] in the continuum
limit in order to reproduce
the results of matrix models.  The first proposal
consisted of an analytic continuation
in the value of the central charge of the Virasoro algebra
\REF\goulian{\rgoulian}[\goulian].  This technique was
further explored in
\REF\kitazawa{\rkitazawa}
\REF\difrankuttwo{\rdifrankuttwo}[\kitazawa,\difrankuttwo].  The second
proposal, closely related to the previous one, was
a generalization of the
Coulomg gas technique
\REF\dotfat{\rdotfat}
[\dotfat]
to the Liouville case in order to include
negative numbers of screening charges
\REF\dotsenko{\rdotsenko} [\dotsenko].  So far
these techniques have only
allowed a direct computation of generic
one-, two,- and three-point functions
and there is no clear procedure known
on how to extend the same ideas to higher
point functions except in some special
 cases.  The second technique was used
in \REF\lagthree{\rlagthree}[\lagthree]
 to clarify some issues concerning the
fusion rules in the presence of $2D$-gravity.

  The basic technical problem
in most of these computations is the evaluation of some integrals
which also appear in the computation of the structure constants of
the minimal conformal models which was done in
[\dotfat].  In the case of $N=1$ Superconformal Fields Theories
\REF\fqsone{\rfqsone}\REF\fqsthree{\rfqsthree}
[\fqsone,\fqstwo,\fqsthree] although there is available a Coulomb gas
formulation
\REF\bkt{\rbkt}[\bkt], there are only partial results with respect to
the structure constants of the operator algebra (see for example
\REF\eich{\reich}
\REF\qiu{\rqiu}
\REF\mussardo{\rmussardo}
[\eich, \qiu, \mussardo]).  The generalization
of the results in [\dotfat] to the minimal $N=1$ models in the
Neveu-Schwarz sector
 has been carried out in
\REF\lagfour{L. Alvarez-Gaum\'e and
 Ph. Zaugg, CERN-TH-6242/91.}[\lagfour].  As
a simple application of the results in this paper we can calculate the
one-, two- and three-point functions of $N=1$ minimal
models coupled
to $2D$-supergravity on surfaces of spherical topology.  The derivation
of the supergravitational dressing and dimensions was carried out in
the light-cone gauge in
\REF\pz{\rpz}[\pz] and in the superconformal gauge
in \REF\dhk{\rdhk}[\dhk].  We follow the
 proposal in [\dotsenko], although
one could equally well extend the ideas
in  [\goulian].  These computations
have some interest because there is as yet
no analogue of the matrix model
formulation for superconformal theories coupled to supergravity
(see nevertheless
\REF\ddk{P. Di Francesco, J. Distler
 and D. Kutasov: PUPT-90-1189, 1990.}[\ddk]).

{\bf 2}. {\it Formulation of the problem}.  In the Coulomb
gas formulation of
the minimal superconformal models [\bkt] the matter
 energy-momentum tensor
is built in terms of a free massless scalar superfield $X (Z)$
$$
T_M(Z)=-{1\over 2}:DX \partial X: +{i\over 2}\alpha_0 D\partial X
\eqn\emtensor
$$
where $Z=(z,\theta)$ represents a point on
the superplane.  It is convenient to
introduce two quantities
\def\ap{\alpha_+}\def\am{\alpha_-}\def\ao{\alpha_0}
$\ap,\am$ satisfying $\ap+\am=\ao\ \ \ \ap \am=-1$.  Then
the central charge and
the screening charges take the form
$$
\hat c=1-2\ao ^2\qquad J_{\pm}=e^{i\alpha_{\pm} X(Z)}
\eqn\candjpm
$$
The background charge $\ao$ changes the dimension
of a super-vertex operator
$e^{i\alpha X}$ from $\alpha ^2/2$ to $\alpha(\alpha-\ao)/2$.
 The minimal
superconformal models are obtained for
 special values of $\ao,\ap,\am$.  Take
two integers $p',p; p'> p$ such that 1)
 $p'-p\equiv 0({\rm mod}\ 2)$, 2) if they
are odd, then they are coprime, or 3)
if both are even then $p'/2,p/2$ are coprime.
In the $N=1$ minimal models we have
$$
\ap=\sqrt{{p'\over p}}\qquad \am=-\sqrt{{p\over p'}}
\qquad \hat c= 1-2{(p'-p)^2\over
pp'}
\eqn\minmod
$$
The primary fields are represented as
 vertex operators in the NS sector
$$
\Psi_{m',m}=e^{i\alpha_{m',m} X}
\qquad m'-m\equiv 0 ({\rm mod}\ \  2)
$$
$$
\alpha_{m',m}={1\over 2}(1-m')\am + {1\over 2}(1-m)\ap
\eqn\nsfield
$$
and in the Ramond sector by
$$
\Psi_{m',m}=\sigma e^{i\alpha_{m',m} X}
\qquad m'-m\equiv 1 ({\rm mod} 2)
\eqn\rfield
$$
Where $\sigma$ is a spin field of the matter sector.
The superconformal dimensions are
$$
h_{m',m}={1\over 8pp'}[(mp'-m'p)^2-(p'-p)^2]+
{1\over 32}(1-(-1)^{m'-m})
\eqn\cdim
$$
We will be exclusively concerned with
correlators of NS fields in this paper.  The
range of $m',m$ is $1\le m' \le p'-1,
1\le m\le p-1,\ \ \ mp'-m'p\ge 0$.

In the super-Liouville sector, after
gauge fixing in the superconformal gauge [\dhk]
we can describe the Liouville part of
the theory locally with an energy-momentum
tensor expressible in terms of a real scalar superfield $\Phi$ :
$$
T_L(Z)=-{1\over 2}D\Phi \partial \Phi +{Q\over 2} D\partial \Phi
\eqn\lemtensor
$$
The superconformal dimension of a
vertex operator in the NS sector
$e^{\beta \Phi}$ is $-\beta (\beta -Q)/2$.  The Liouville
background charge
is determined by $\ao$ and the ghost contributions to be
$$
Q^2={9-\hat c\over 2}=4+\ao^2
\eqn\qasao
$$
The dressing of a NS field $e^{i\alpha X}$ is given by
$$
e^{i\alpha X}e^{\beta \Phi},
\qquad {1\over 2}\alpha(\alpha-\ao)-{1\over
2}\beta(\beta-Q)={1\over2}
\eqn\dressing
$$
making the dressed field into
 a $(1/2,1/2)$-form which can
 be integrated over the
surface without any reference to the
 conformal factor of the background metric.  The
two solutions to the quadratic equation in \dressing\  are
$$
\beta={Q\over 2}\pm|\alpha-{\ao\over 2}|
\eqn\soldressing
$$
The microscopic branch (see [\seiberg] for details)
is obtained by choosing the
minus sign.  With this sign the superconformal
dimension agrees with the classical dimension
in the classical limit
$\hat c\rightarrow -\infty$.  For \nsfield\ we obtain
$$
\beta_{m',m}={p+p'-|mp'-m'p|\over 2\sqrt{pp'}}
\eqn\gravdim
$$
As with the conformal case we can
introduce two quantities, $\beta_{\pm}$ satisfying
\def\bp{\beta_+}\def\bm{\beta_-}
$$
\bp+\bm =Q\qquad \bp \bm=1
\eqn\bpbm
$$
They are related to $\ap,\am$ by
$$
\ap=\bp\qquad\am=-\bm
\eqn\abrel
$$
It is useful to notice that in super-Liouville
theory the cosmological constant
couples to the operators $e^{\bm \Phi}$.  Formally
 we can introduce two screening
charges in the Liouville sector
$$
U_{\pm}=e^{\beta_{\pm}\Phi}
\eqn\lscreening
$$
If we are interested in general
 properties of $n$-point correlators we first note that
the area constraint in the conformal
 case is replaced here by a "length" constraint
because we are counting volumes of $N=1$
supersurfaces.  For arbitrary values of
$p',p$ the field of lowest superconformal
 dimension is not necessarily the identity.
Let \def\mmin{\Psi_{{\rm min}}} $\mmin$ be
 the field of lowest dimension, and let
\def\bmin{\beta_{{\rm min}}}$\bmin$ be the
 corresponding dressing exponent.  The
cosmological constant $\mu $ should be taken
 as the coefficient of the operator
$\mmin e^{\bmin \Phi}$.  If we concentrate for
 the time being on the super-Liouville
contribution to the $n$-point function, we can
obtain a useful expression for
$\langle \prod_i e^{\beta_i\Phi(Z_i)}\rangle$ by
 following step by step the arguments
in the conformal case ([12], see also
\REF\berkleb{\rberkleb}[\berkleb],[5]).  First we
introduce a $\delta$-function
constraint fixing the length and integrate over
lengths from $0$ to $\infty$.
Second, we separate the constant piece from
 $\Phi(Z_i)$ to explicitly solve the
$\delta$-function, $\Phi(Z_i)=\phi_0+\tilde {\Phi}(Z_i)$.  For
 a surface of genus
$h$, the result is
$$
\langle \prod_i e^{\beta_i\Phi(Z_i)}\rangle=
{1\over \bmin}\mu^s\Gamma(-s)\langle\tilde L^s \prod_i
e^{\beta_i\tilde{\Phi}(Z_i)}\rangle _{S_L^{(0)}(\tilde{\Phi})}
$$
$$
s=-{1\over \bmin}(\sum_i\beta_i-Q(1-h))
\eqn\goulli
$$
and
$$
\tilde L=\int d^2Z \hat E e^{\bmin \tilde \Phi}\mmin
\eqn\ltilde
$$
$\hat E$ is the reference zweibein
 and the integration over supermoduli
parameters is not explicitly
 exhibited.  The expectation value in
\goulli\ is computed in terms of
 the free super-Liouville action with a
term representing the
background charge $Q\hat R \tilde \Phi$ ($\hat R$ is the curvature
associated to the superframe $\hat E$; see
 [\dhk] for details).  Note parenthetically
that we can read off from the $\mu$-dependence
 in \goulli\ the string susceptibility
and the gravitational dimensions of the fields
in the theory.  If $s$ were
a positive integer
we could easily evaluate \goulli\ using
 free field techniques.  This is the method
proposed in [\goulian].  One first computes
 the integral for $s$ an integer and then
one analytically continues to arbitrary values.  This
prescription could also be applied to our case.
We choose however to follow [\dotsenko].  The idea
 is to treat the Liouville
and matter sectors
in the same way using the Coulomb gas formulation.  We should
make a few remarks before
we proceed.  When the matter theory is unitary, the field
of lowest dimension is the identity
 operator $\mmin=1$. In the computation we will
describe below for the general case, we
 take the cosmological constant by definition
to be the coefficient of the dressed
identity operator.  In the non-unitary case this
corresponds to a fine tuning of the
coupling of all the operators of negative
dimension.  The derivation of (15) goes through
again, with the only
change that $\bmin$ is replaced by $\beta_-$.  We can
introduce vertex operators similar to (4):
$$
\eqalign{ L_{m',m}&=e^{\beta_{m',m}\Phi}
\cr \beta_{m',m}&={1-m'\over 2}\beta_-+
{1-m\over 2}\beta_+}
\eqn\lvertex
$$
Since the dressing exponent of (4) is
$$
\beta(\alpha_{m',m})={Q\over 2}-
{1\over 2}\beta_-|m'-\rho m|\qquad \rho={\beta_+\over
\beta_-}
$$
we can represent $\beta(\alpha_{m',m})$
 in terms of $\beta_{m',m}$.  There are two
cases to distinguish:
$$
\eqalign{m'>\rho m,\qquad \beta(\alpha_{m',m})
&={Q\over 2}-{1\over 2}\beta_-(m'-\rho
m)=\beta_{m',-m} \cr
 m'<\rho m,\qquad \beta(\alpha_{m',m})&={Q\over 2}-{1\over
2}\beta_-(-m'+\rho m)=\beta_{-m',m}}
\eqn\gdressing
$$
We can assign a chirality  $\chi$ to a
primary field $\Psi_{m',m}$.  $\chi=1$
if $m'>\rho m$ ; $\chi=-1$ if $m'<\rho m$.  To guarantee
 the choice of the microscopic
branch in (10) there are two possible dressings
$$
\eqalign{A^+_{m',m}&=\Psi^+_{m',m}L_{m',-m}\qquad \chi =1\cr
A^-_{m',m}&=\Psi^-_{m',m}L_{-m',m}\qquad \chi =-1}
\eqn\afields
$$
The dressed fields have dimensions
$(1/2,1/2)$.  Ignoring for the moment chirality
labels, we are interested in
$$
\langle A_S A_N A_M \rangle=\int
\prod_1^3d^2Z_i{\hat E}\langle A_S(Z_3) A_N(Z_2)
A_M(Z_1)\rangle
\eqn\threecorr
$$
using $SL(2|1)$ invariance we can
fix $Z_1=(\infty,\infty\eta), Z_2=(1,0),Z_3=(0,0)$.
Dividing out the $SL(2|1)$ volume
we are left with a single integration over the odd
variable $\eta$:
$$
\eqalign{\langle A_S A_N A_M \rangle=
\int d^2\eta \langle A_S(0,0) A_N(1,0)
A_M(\infty,\infty\eta)\rangle= \cr
\int d^2\eta \langle  \Psi_S(0,0)
\Psi_N(1,0) \Psi_M(\infty,\infty\eta)\rangle
\langle L_S(0,0) L_N(1,0) L_M(\infty,\infty\eta)\rangle}
\eqn\intone
$$
The matter correlation function is evaluated
using the physical structure constants
[\lagfour]
$$
\langle  \Psi_S(0,0) \Psi_N(1,0) \Psi_M(\infty,\infty\eta)\rangle =
D^{{\rm even}}_{SNM} + D^{{\rm odd}}_{SNM}|\eta|^2
\eqn\mattercorr
$$
The contribution will come from $D$ even or
odd depending on the number of screening
charges needed in the evaluation of
\mattercorr\ in the Coulomb gas formulation.  For
$\langle A_S A_N A_M \rangle$ to be
non-vanishing, the matter and Liouville parts
must have opposite Grassmann parity.  This will be verified later by counting
screenings in both cases.  Using capital
 letters to label pairs of indices
$(M=(m',m)\ $,etc), the physical structure
 constants $D$ are defined in terms of
the symmetric structure constants in the
Coulomb gas definition of the correlators
$$
C_{SNM}=\langle \Psi_S \Psi_N \Psi_M ({\rm screenings})\rangle
\eqn\symmconst
$$
Defining the conjugate fields $\Psi_{\overline S}\equiv \Psi_{-S},
(\overline s',\overline s)=(-s',-s)$,
the asymmetric structure constants are defined
according to
$$
C_{SN}^M=
\langle \Psi_S \Psi_N \Psi_{\overline M} ({\rm screenings})\rangle
\eqn\nonsymmconst
$$
In analogy with [\dotfat] it is shown in
[\lagfour] for the superconformal case that
the physical structure constants $D_{SNM}$
are simply related to $C_{SNM}$.  There
are four equivalent ways of writing $D$ in terms of $C$:
$$
\eqalign{D_{SNM}=&(C^M_{SN} C^S_{MN}a_N)^{1/2}\cr
=&(a_N a_S a_M^{-1})^{1/2} C^M_{SN}\cr
=&(a_N a_M a_S^{-1})^{1/2} C^S_{MN}\cr
=&\Omega^{-1} (a_N a_S a_M)^{1/2} C_{SNM}\cr
a_S=&(C^1_{SS})^{-1}\qquad \Omega = -{\rho \over 4}\Delta({\rho-1\over
2})\Delta({\rho'+1\over 2})}
\eqn\dasc
$$
  Let $Q=(q',q)$ be the number of screenings
in the matter sector.  $q'$ (resp. $q$) counts
 the number of $J_-$ (resp. $J_+$)
screening charges in the correlator.  The parity
 of $q'+q$ is the same for the four
representations in \dasc .  The number of screenings
 in the Liouville sector depends
on $q',q$.  Consider first the case with
 all chiralities $\chi=-1$, $\langle - -
-\rangle$:
$$
\langle L_{-s',s}(0,0)  L_{-m',m}(1,0)  L_{-n',n}
(\infty,\infty \eta)\rangle
\eqn\lthree
$$
The Liouville screenings are
$$
\eqalign{q'_L=&{-s'-n'-m'-1\over 2}=-q'-1\cr q_L=&{s+n+m-1\over 2}=q}
$$
hence the parity of the Liouville sector
$$
q'_L+q_L=-q'-1+q\equiv q'+q-1 \ ({\rm mod}\  2)
$$
is opposite to the matter parity as
required for the non-vanishing of three-point
correlators.  For other chiralities
it is easy to show that the same conclusion
holds.  The three-point functions can
be expressed as products of ratios of
$\Gamma$-functions, with the products
ranging up to $q-1$ or $q'-1$.  For the matter
sector these numbers are positive.  In the
 Liouville case however one of them is
negative.  The analytic continuation advocated
 in [\dotsenko] is to use the definition
$$
\prod_{i=1}^{-l'-1}f(i)=\prod_{i=0}^{l'}{1\over f(-i)}\qquad l'>0
\eqn\dotanal
$$
This together with the results of [\lagfour]
is all we need to write the explicit
form of the three-point correlators.  Before
we write the results, we have to
determine the $\mu$-dependence of
the correlators.  The coefficient of
$U_-$ in the action is $\mu$.  However,
the coefficient of $U_+$ in the Liouville
action is taken as in [\dotsenko] to be
 determined by its gravitational dimension,
and it is given by $\mu^{\rho}$.  This
 guarantees that the power of $\mu$
in front of the correlators is exactly $s$ as defined in (15).

{\bf 3.} {\it Computations}.  There are four cases
 depending on the chirality:
$\langle - - - \rangle$ , $\langle + - - \rangle$,
 $\langle + + - \rangle$,
$\langle + + + \rangle$.  The computations
are very similar in all four cases.  Although
the  matter and Liouville correlators are
 separately very cumbersome, when both of
them are put together almost everything
 cancels leaving only
a set of terms similar to Polyakov leg factors
\REF\polyaleg{A.M. Polyakov:  Mod. Phys. Lett. {\bf A 6}(1991) 635.}
[\polyaleg].  Depending on the
 chirality of the matter fields the cancellation
is made more evident by choosing
one of the representations \dasc.  We list now
each chirality case together with
the number of screenings involved
in the Liouville and matter
sectors.  Take $k',k$ to be the number of $-,+$
screening operators in the Liouville sector,
and $l',l$ in the matter sector.  The
four cases to be considered are

i).  $\langle - - - \rangle$
$$
\eqalign{k'=&{1\over 2}(-s'-n'-m'-1)=-l_1'-1 \cr
k=&{1\over 2}(s+n+m-1)=l_1}
\qquad \eqalign{l_1'=&{1\over 2}(s'+n'+m'-1)\cr
l_1=&{1\over 2}(s+m+n-1)}
\eqn\casei
$$
and in performing the cancellation it is best
 to take $D_{SNM}\sim C_{SNM}$.

ii).  $\langle + + - \rangle$
$$
\eqalign{k'=&{1\over 2}(s'+n'-m'-1)=l_2' \cr
k=&{1\over 2}(-s-n+m-1)=-l_2-1}
\qquad \eqalign{l_2'=&{1\over 2}(s'+n'-m'-1) \cr
l_2=&{1\over 2}(s+m+n-1)}
\eqn\caseii
$$
and we take $D_{SNM}\sim C^M_{SN}$.

iii).  $\langle + - - \rangle$
$$
\eqalign{k'=&{1\over 2}(s'-n'-m'-1)=-l_3'-1 \cr
k=&{1\over 2}(-s+n+m-1)=l_3}
\qquad \eqalign{ l_3'=&{1\over 2}(-s'+n'+m'-1) \cr
l_3=&{1\over 2}(-s+m+n-1)}
\eqn\casei
$$
and $D_{SNM}\sim C^S_{NM}$.

iv).  $\langle + + +\rangle$
$$
\eqalign{k'=&{1\over 2}(s'+n'+m'-1)=l_1' \cr
k=&{1\over 2}(-s-n-m-1)=-l_1-1}
\eqn\caseiv
$$
and $D_{SNM}\sim C_{SNM}$.

The matter three-point function up to
irrelevant constants can be represented as a
suface integral
$$
\eqalign{ \lim_{R \rightarrow \infty} R^{4\Delta (\alpha_M)} &\langle
V_{\alpha_M}(R,R\eta) V_{\alpha_N}(1,0) V_{\alpha_S}(0,0)
\int \prod_1^{l'} d^2Z'_i\; V_\alm (Z'_i,\overline{Z}'_i)
\int \prod_1^l d^2Z_i\; V_\alp(Z_i,\overline{Z}_i)
\rangle \cr
  = \lim_{R \rightarrow \infty} & R^{4\Delta (\alpha_M)}
\int \prod_1^{l'} d^2Z'_i \int \prod_1^l d^2Z_i \;
|R-1|^{2\alpha_M \alpha_N}  |R|^{2\alpha_M \alpha_S}
|R-z_i-R\eta\theta_i|^{2\alpha_M \alp} \cr
  & |R-z'_i-R\eta\theta'_i|^{2\alpha_M \alm}
\prod_{i=1}^{l'}|1-z'_i|^{2\alpha_N\alm} |z'_i|^{2\alpha_S\alm}
\prod_{i<j}^{l'} |z'_i -z'_j -\theta'_i \theta'_j|^{2\alm^2} \cr
  & \qquad \prod_{i=1}^l |1-z_i|^{2\alpha_N\alp} |z_i|^{2\alpha_S\alp}
\prod_{i<j}^l |z_i -z_j -\theta_i \theta_j|^{2\alp^2}
\prod_{i,j}^{l,l'} |z_i-z'_j-\theta_i \theta'_j|^{-2} }
\eqn\mattint
$$
with the screening condition
$$
\alpha_M+\alpha_N+\alpha_S +l\ap + l'\am=\ao
\qquad \rho=\ap^2={\rho '}^{-1}=\am^{-2}
$$
$$
\alpha_S=\alpha_{s',s},...
$$
Simple manipulations in \mattint\  yield
$$
\eqalign{ \int \prod_1^{l'} d^2Z'_i \int & \prod_1^l d^2Z_i \;\xi\;
\prod_{i=1}^{l'} |z'_i|^{2a'} |1-z'_i|^{2b'} \prod_{i<j}^{l'}
|z'_i -z'_j -\theta'_i \theta'_j|^{2\rho'} \cr
  & \prod_{i=1}^l |z_i|^{2a} |1-z_i|^{2b}
\prod_{i<j}^l |z_i -z_j -\theta_i \theta_j|^{2\rho} \prod_{i,j}^{l,l'}
|z_i-z'_j-\theta_i \theta'_j|^{-2} }
\eqn\mattintbis
$$
where
$$
\eqalign{\xi=&|1- \alpha_M \ap\sum_i
\eta\theta_i- \alpha_M\am\sum_i\eta\theta_i'|^2\cr
a'=&\am\alpha_{s',s}\qquad a=\ap \alpha_{s',s} \cr
b'=&\am\alpha_{n',n}\qquad b=\ap \alpha_{n',n} \cr
c'=&\am\alpha_{m',m}\qquad c=\ap \alpha_{m',m} }
$$
The integral \mattintbis\ is given by the structure constants
$C_{SNM}=C^{-M}_{SN}$ which can be found in [\lagfour]:
$$
\eqalign{ (\const)\rho^{2l l'} & \left({\rho' \over 2}
\right)^{2 M'_{l'}} \left({\rho \over 2}\right)^{2 M_l+ 2M'_{l'+1}} \cr
  \prod_1^l & \Delta(-l' +\rop i -M_i)
\prod_1^{l'} \Delta(\ropp i -M'_i) \cr
  \prod_0^{l-1} & \Delta(1+a+\rom i + M_i) \Delta(1+b+\rom i + M_i) \cr
  & \quad \Delta(-a-b-\rho(l-1)+l'+\rom i + M_i) \cr
  \prod_0^{l'-1} & \Delta(1+a'+\ropm i + M'_i) \Delta(1+b'+\ropm i +
M'_i)\cr
  & \quad\Delta(-a'-b'-\rho'(l'-1)+l+\ropm i + M'_i) }
\eqn\gfinal
$$
where
$$
M'_i=[{i\over 2}]\qquad M_i=-l'+[{l'+i\over 2}]
$$
with $[x]=$ integer part of $x$ and $\Delta(x)=
\Gamma(x)/\Gamma(1-x)$.  Furthermore
from the screening conditions we have
$$
\eqalign{-a'-b'-\rho'(l'-1)+l=1+c'\cr -a-b-\rho(l-1)+l'=1+c}
$$
The same techniques apply to the Liouville part
$$
\eqalign{ \lim_{R \rightarrow \infty} R^{4\Delta(\beta_M)} &\langle
V_{\beta_M}(R,R\eta) V_{\beta_N}(1,0) V_{\beta_S}(0,0)
\int \prod_1^{k'} d^2Z'_i \; V_\bem (Z'_i,\overline{Z}'_i)
\int \prod_1^k d^2Z_i \; V_\bep(Z_i,\overline{Z}_i)
\rangle \cr
  = & \int \prod_1^{k'} d^2Z'_i \int \prod_1^k d^2Z_i \;\xi\;
\prod_{i=1}^{k'} |z'_i|^{2a'} |1-z'_i|^{2b'} \prod_{i<j}^{k'}
|z'_i -z'_j -\theta'_i \theta'_j|^{-2\rho'} \cr
  &\quad \prod_{i=1}^k |z_i|^{2a} |1-z_i|^{2b}
\prod_{i<j}^k |z_i -z_j -\theta_i \theta_j|^{-2\rho} \prod_{i,j}^{k,k'}
|z_i-z'_j-\theta_i \theta'_j|^{-2} }
\eqn\liouint
$$
where now
$$
\eqalign{a'=&-\beta_-\beta_S\qquad a=-\beta_+\beta_S \cr
b'=&-\beta_-\beta_N\qquad b=-\beta_+\beta_N \cr
c'=&-\beta_-\beta_M\qquad c=-\beta_+\beta_M }
$$
and $k',k$ depend on the chiralities.  This
 integral is as in the matter sector
except for the fact that $\rho,\rho'$ are
 changed into $-\rho,-\rho'$.  We
finally have
$$
\eqalign{  (\const) (-\rho)^{2k k'} & \left(-{\rho' \over 2}
\right)^{2 M'_{k'}} \left(-{\rho \over 2}\right)^{2 N_k+ 2M'_{k'+1}} \cr
  \prod_1^k & \Delta(-k' -\rom i -N_i)
\prod_1^{k'} \Delta(-\ropm i -M'_i) \cr
  \prod_0^{k-1} & \Delta(1+a-\rop i + N_i) \Delta(1+b-\rop i + N_i) \cr
  & \quad \Delta(-a-b+\rho(k-1)+k'-\rop i + N_i) \cr
  \prod_0^{k'-1} & \Delta(1+a'-\ropp i + M'_i) \Delta(1+b'-\ropp i +
M'_i)\cr
  & \quad\Delta(-a'-b'+\rho'(k'-1)+k-\ropp i + M'_i) }
\eqn\lfinal
$$
now
$$
\eqalign{M'_i=&[{i\over 2}]\qquad N_i=-k'+[{k'+i\over 2}] \cr
1+c'=&-a'-b'+\rho '(k'-1)+k \cr
1+c=&-a-b+\rho (k-1) +k'}
$$
Using \dotanal\  and standard properties
of $\Gamma$-functions we obtain after some
tedious computations

i). $ \langle - - -\rangle$
$$
\eqalign{\langle \Psi^-_{m',m}(\infty,\infty\eta)
 \Psi^-_{n',n}(1,0)\Psi^-_{s',s}(0,0)\rangle
\langle L_{-m',m}(\infty,\infty\eta) L_{-n',n}(1,0)
 L_{-s',s}(0,0)\rangle \cr
=\mu^s C_1\Delta(1-{s\over 2}+{s'\over 2}\rho')
\Delta(1-{n\over 2}+{n'\over 2}\rho') \Delta(1-{m\over 2}
+{m'\over 2}\rho')}
\eqn\resulti
$$
ii).   $ \langle + + -\rangle$
$$
=\mu^s C_2\Delta(1-{s'\over 2}+{s\over 2}\rho)
\Delta(1-{n'\over 2}+{n\over 2}\rho) \Delta(1+{m'\over 2}+
{m\over 2}\rho)
\eqn\resultii
$$
iii).   $ \langle + - -\rangle$
$$
=\mu^s C_3\Delta(1+{s\over 2}-{s'\over 2}\rho')
\Delta(1-{n\over 2}+{n'\over 2}\rho') \Delta(1-{m\over 2}
+{m'\over 2}\rho')
\eqn\resultiii
$$
iv).   $ \langle + + +\rangle$
$$
=\mu^s C_4\Delta(1-{s'\over 2}+{s\over 2}\rho)
\Delta(1-{n'\over 2}+{n\over 2}\rho) \Delta(1-
{m'\over 2}+{m\over 2}\rho)
\eqn\resultiv
$$
There is a common infinite factor in $C_i$ which
 can be absorbed in the normalization
of the correlation function.  Recall also
 that $s$ is given by (15) but with
$\bmin$ replaced by $\beta_-$.  As a particular
 application, and to compare with
the results in the conformal case
[\goulian, \dotsenko] we consider the
unitary case $p'=p+2$ and diagonal
fields $A_{m,m}$.  It is convenient to
evaluate a combination of correlators
 where the dependence on $\mu$ cancels
$$
{\langle A_{m_1,m_1}  A_{m_2,m_2}  A_{m_3,m_3} \rangle ^2 Z
\over \langle A_{m_1,m_1}  A_{m_1,m_1}\rangle
\langle A_{m_2,m_2}A_{m_2,m_2}\rangle
 \langle A_{m_3,m_3}A_{m_3,m_3}\rangle}=
{m_1m_2m_3\over (1+\rho)\rho(\rho -1)}
\eqn\compare
$$
In the case when $p'=p+2 =2(n+1)$ these
correlators cannot be distinguished from
the same diagonal correlators in the $(n+1,n)$
 conformal theory.  In this case
not only the string susceptibility and the
 gravitational dimensions coincide, but
also the special combination \compare\  of
 three-point functions.  The arguments of the
$\Gamma$-functions in \resulti-\resultiv\
can be written in a simpler form if we
combine the quantities $\alpha,\beta$ of
 a dressed vertex operator
$e^{i\alpha X}e^{\beta \Phi}$ into a Minkowskian
 two-dimensional vector
$p=(\beta,\alpha)$.  Defining the background
 two-vector $b=(Q,\ao)$, the dressing
(on-shell) condition takes the form $(p-b/2)^2=0$,
 where for any two-vector
$p^2=\beta^2-\alpha^2$.  Then the
argument $1-m/2+m'\rho/2$ appearing for
example in \resulti\  can be written
 as ${1\over 2}+{p^2\over 2}$.  Finally the
singularities of the factors in
\resulti-\resultiv\  appear in the boundary of the
Kac table for values of $(m',m)=(0,m)$, with
$m$ even in analogy with the
conformal case [\polyaleg].  These
states should very likely have an interpretation
as boundary operators as suggested
for the conformal case in
\REF\mms{E. Martinec, G. Moore and N. Seiberg,RU-14-91,YCTP-P10-91,
EFI-91-14.}[\mms].

We find it rather intriguing that in
the case when $p'=p+2$ is even (the case when
the theory has a supersymmetric ground state
 before coupling to gravity), the zero-,
one-, two- and three-point functions in the NS
sector of the theory cannot be
distinguished from the conformal case $(1+{p\over 2},{p\over 2})$.

While this paper was being typed we
received three papers where similar issues are
addressed
\REF\abdalla{E. Abdalla, M.C.B. Abdalla,
D. Dalwazi, and K. Harada, Print-91-0351 (Sao
Paulo).}
\REF\difrankutfour{P. Di Francesco and D. Kutasov, PUPT-1276.}
\REF\dhoker{K. Aoki and E. D'Hoker< UCLA-91-TEP-33.}
[\abdalla,\difrankutfour,\dhoker].

ACKNOWLEDGEMENTS.  We would like to
thank M. Ruiz-Altaba for many discussions during
the course of this work.

\endpage
\refout
\end